\documentclass[12pt]{iopart}

\usepackage{graphicx,epsfig}
\usepackage{longtable}
\usepackage{color}
\usepackage[sort&compress,numbers]{natbib}
\usepackage{doi}
\usepackage{hyperref}

\usepackage{bm}
\usepackage{capt-of}
\usepackage{relsize}
\usepackage{xcolor}
\usepackage{amssymb}
\usepackage[normalem]{ulem}
\usepackage{mathptmx}

\def\al{\alpha}
\def\be{\beta}
\def\ga{\gamma}

\def\Ga{\Gamma}
\def\De{\Delta}

\def\beq{\begin{eqnarray}}
\def\eeq{\end{eqnarray}}

\def\uGeVct{\mbox{GeV}/\mbox{c}^2}

\def\numu{\nu_\mu}
\def\numubar{\bar{\nu}_\mu}

\def\piz{\pi^\circ}
\def\pip{\pi^{+}}
\def\pim{\pi^{-}}

\begin{document}

\title{PYTHIA hadronization process tuning in GENIE neutrino interaction generator}

\author{Teppei Katori and Shivesh Mandalia}

\address{School of Physics and Astronomy, Queen Mary University of London, London E1 4NS, UK}
\ead{t.katori@qmul.ac.uk}
\vspace{10pt}
\begin{indented}
\item[]July 2015
\end{indented}

\begin{abstract}
Next generation neutrino oscillation experiments utilize details 
of hadronic final states to improve the precision of neutrino interaction measurements.  
The hadronic system was often neglected or poorly modeled in the past, 
but they have significant effects on high precision neutrino oscillation and cross-section measurements. 
Among the physics of hadronic systems in neutrino interactions, 
the hadronization model controls multiplicities and kinematics
of final state hadrons from the primary interaction vertex. 
For relatively high invariant mass events, 
many neutrino experiments rely on the PYTHIA program. 
Here, we show a possible improvement of this process in neutrino event generators,  
by utilizing expertise from the HERMES experiment. 
Finally, we estimate the impact on the systematics of hadronization models 
for neutrino mass hierarchy analysis using atmospheric neutrinos such as the PINGU experiment. 
\end{abstract}

\pacs{11.80.Cr,13.15.+g,14.60.Lm,14.60.Pq,25.30.Fj,25.30.Pt,95.85.Ry}
%
\vspace{2pc}
\noindent{\it Keywords}: neutrino cross section, hadronization, PYTHIA, GENIE
%
%
%
%

\section{Introduction, future long baseline neutrino oscillation experiments}

Neutrino oscillations are function of the neutrino baseline L,
and neutrino energy $E_\nu$.
By optimizing factors such as cost, neutrino production, and detection processes, 
typical oscillation experiments often choose L$\sim 100-1000$~km and $E_\nu~\sim 1-10$~GeV. 
Some examples include T2K~\cite{T2K_osc},
which uses a 600 MeV off-axis J-PARC neutrino beam~\cite{T2K_flux},
and NOvA~\cite{NOvA},
which uses a 2~GeV off-axis NuMI beam~\cite{NuMI}.  
Their flux peaks are tuned to quasielastic and resonance dominant regions 
in order to perform neutrino oscillation measurements. 
Therefore, in the past few years,
the neutrino interaction community has spent 
a significant amount of time trying to understand the physics 
at these energy regions~\cite{Gallagher_review,Morfin,Hayato,Zeller_INT},
especially after the discovery of the importance of
two-body current in neutrino physics~\cite{Teppei_CCQE,Martini_first,Nieves_first}.

Although off-axis beams from J-PARC and NuMI are tuned to narrow 600~MeV and 2~GeV peaks, 
off-axis neutrino beams made from wide-band decay-in-flight mesons have long high-energy tails, 
and the contribution from large $W$ interactions is always present. 
For example, multi-pion production processes contribute significant amounts in 
single pion production measurements at T2K~\cite{Connolly},
and NOvA~\cite{NOvA} uses hadronic shower information to reconstruct the neutrino energy. 
On top of that, future long baseline oscillation experiments, such as PINGU~\cite{PINGU}, ORCA~\cite{ORCA},
Hyper-Kamiokande~\cite{HyperK}, and DUNE~\cite{LBNE}
are trying to use hadronic information from the atmospheric neutrino interactions around 2-20 GeV.
Therefore correct modeling of hadronization processes is an important subject
for current and future neutrino oscillation experiments.

In this paper, we study how the hadronization model based on PYTHIA can be improved by utilizing
HERMES expertise for current and future neutrino oscillation experiments. 
In Sec.~\ref{sec:pythia} and Sec.~\ref{sec:genie}, 
we introduce the PYTHIA hadronization simulator and the GENIE neutrino interaction generator,
two of the most important tools in our studies.
Then we briefly discuss the HERMES experiment in Sec.~\ref{sec:hermes}.
Our main results are described in
Sec.~\ref{sec:avmulch},~\ref{sec:xf},~\ref{sec:avmulpiz}, and~\ref{sec:topo}.
Finally, in Sec.~\ref{sec:pingu},
we demonstrate the impact of hadronization models for the PINGU experiment.

\section{PYTHIA, the standard hadronization model\label{sec:pythia}}

\begin{figure}[tb]
  \includegraphics[width=0.35\textheight]{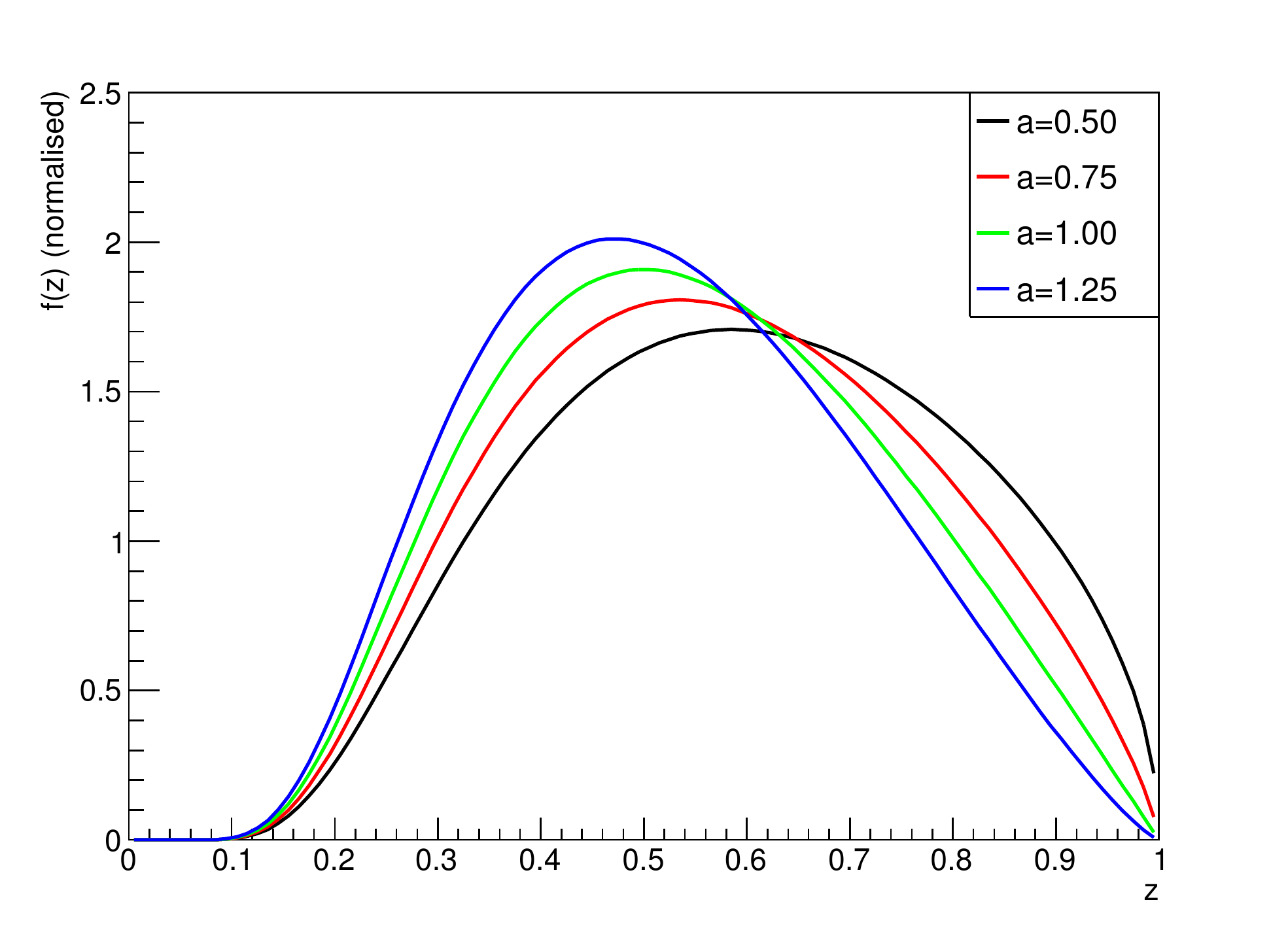}
  \includegraphics[width=0.35\textheight]{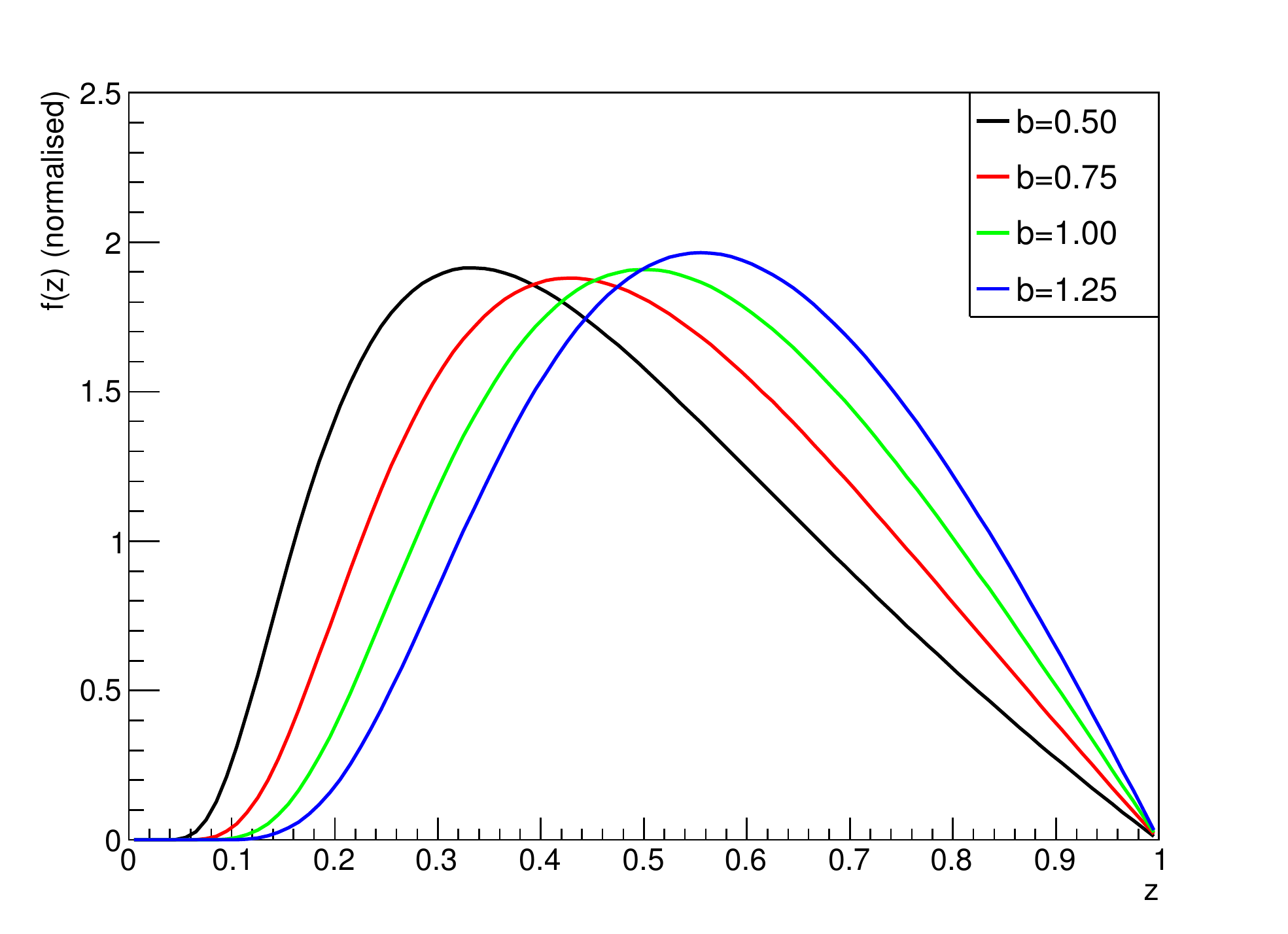}
\caption{\label{fig:lundfunc}
  (color online)
  The Lund symmetric fragmentation function
  (Eq.~\ref{eq:lundff}) for different values of
  Lund $a$ and Lund $b$. The parameter is altered while keeping all
  other variables fixed.
}
\end{figure}

The PYTHIA Monte Carlo (MC) generator~\cite{PYTHIA6,PYTHIA8}
is regarded as one of the standard hadronization tool for high energy physics experiments.  
Fragmentation in PYTHIA is described by the Lund string fragmentation model, 
which is a model based on the dynamics of one-dimensional 
relativistic strings that are stretched between colored partons. 
These strings represent the color flux and in particular, 
are subject to a linear confinement potential. 
The hadronization process is described by break-ups in the strings through 
the production of a new quark-antiquark pairs. 
An iterative approach is used to perform the fragmentation as each break up is causally disconnected. 
The production rate of the created $q\bar{q}$ pair is determined using the tunneling mechanism, 
which leads to a Gaussian spectrum of the transverse momentum, $p_\perp^2 (=p_x^2+p_y^2)$, 
for the produced hadron. 
The fraction of $E+p_z$ taken by the produced hadron is given by the variable $z$, 
defined by the hadron energy $E$ and energy transfer $\nu$ ($z=E/\nu$). 
An associated fragmentation function $f(z)$ gives the probability that a given $z$ is chosen. 
The simplified Lund symmetric fragmentation function is given by,
\beq 
f(z)\propto z^{-1}(1-z)^{a}\cdot exp\left(\frac{-bm_\perp^2}{z}\right)~.
\label{eq:lundff}
\eeq
Here, $m_\perp^2$ is the transverse mass of the hadron ($m_\perp^2\equiv m^2+p_\perp^2$). 
The Gaussian term describes quantum tunneling in the transverse direction, 
and tunable  ``Lund $a$'' and ``Lund $b$'' parameters decide the longitudinal distribution of energy. 
Thus, these two parameters mainly decide how to distribute available energy to the produced hadrons. 
Fig.~\ref{fig:lundfunc} shows the Lund symmetric function.
Larger Lund $a$ and smaller Lund $b$ parameters shift
the fragmentation function to a lower $z$ region. 
The values of these parameters are obtained from the shapes of the measured fragmentation functions, 
and default values of Lund $a$ and Lund $b$ in PYTHIA6.3 are 0.3 and 0.58~$\uGeVct$ respectively.

\section{GENIE AGKY model\label{sec:genie}}
GENIE is a ROOT-based neutrino interaction MC generator~\cite{GENIE}. 
In the few-GeV energy region which are particularly important in oscillation experiments,
GENIE employs a new hadronization model 
called the AGKY model~\cite{AGKY,Tingjun}.

The AGKY model is split into two parts.
At lower energy regions where PYTHIA hadronization models deteriorate, 
a phenomenological description based on the Koba-Nielson-Olesen (KNO) scaling law is used~\cite{KNO}.
First, averaged charged hadron multiplicity data are fit
to a function of invariant mass squared, $W^2$,
in order to extract the parameters $a_{ch}$ and $b_{ch}$, 
\beq
\left<n_{ch}\right>=a_{ch}+b_{ch}\cdot logW^2~,~
\label{eq:abW2}
\eeq
then, the total averaged hadron multiplicity is deduced
to be $\left<n_{tot}\right>=1.5\left<n_{ch}\right>$.
In this way, averaged hadron multiplicity is assigned for any interaction. 
To simulate the actual hadron multiplicity for each interaction, the KNO scaling law is used. 
The KNO scaling law relates the dispersion of hadron multiplicity at different invariant masses 
with a universal scaling function $f(n/\left< n\right>)$,
\beq
 \left< n \right> \times P(n) = f\left(\frac{n}{\left< n\right>}\right)
\label{eq:KNO}
 \eeq
where $\left<n\right>$ is the averaged hadron multiplicity and $P(n)$ is the probability of generating $n$ hadrons. 
The scaling function is parameterized by the Levy function,
\beq
L(z,c)= \frac{2 e^{-c}c^{cz + 1}}{\Ga(cz+1)}~,~
\label{eq:Levy}
\eeq
$z=n/\left< n\right>$ and an input parameter $c$. 
The input parameter is used to tune the function to agree with data,
which is mainly taken from the Fermilab 15-foot bubble chamber experiment~\cite{Zieminska}.
Although more recent hadron multiplicity data are available from CHORUS~\cite{CHORUS},
the heavy target data require more sophisticated 
final state interaction models to access to the primary hadron multiplicity information, 
and we do not take these into account in this article. 

At higher energy regions the AGKY model gradually transitions from the KNO scaling-based model to PYTHIA discussed previously. 
A transition window based on the value of the invariant hadronic mass $W$ is used, 
over which the fraction of events hadronized using the PYTHIA(KNO) model increases(decreases) linearly. 
The default values used in the AGKY model are 
\begin{itemize}
\item $W<2.3~\uGeVct$, KNO scaling-based model only region, 
\item $2.3~\uGeVct<W<3.0~\uGeVct$, transition region, and 
\item $3.0~\uGeVct<W$, PYTHIA only region.
\end{itemize}

\begin{figure}[tb]
\includegraphics[width=0.6\textheight]{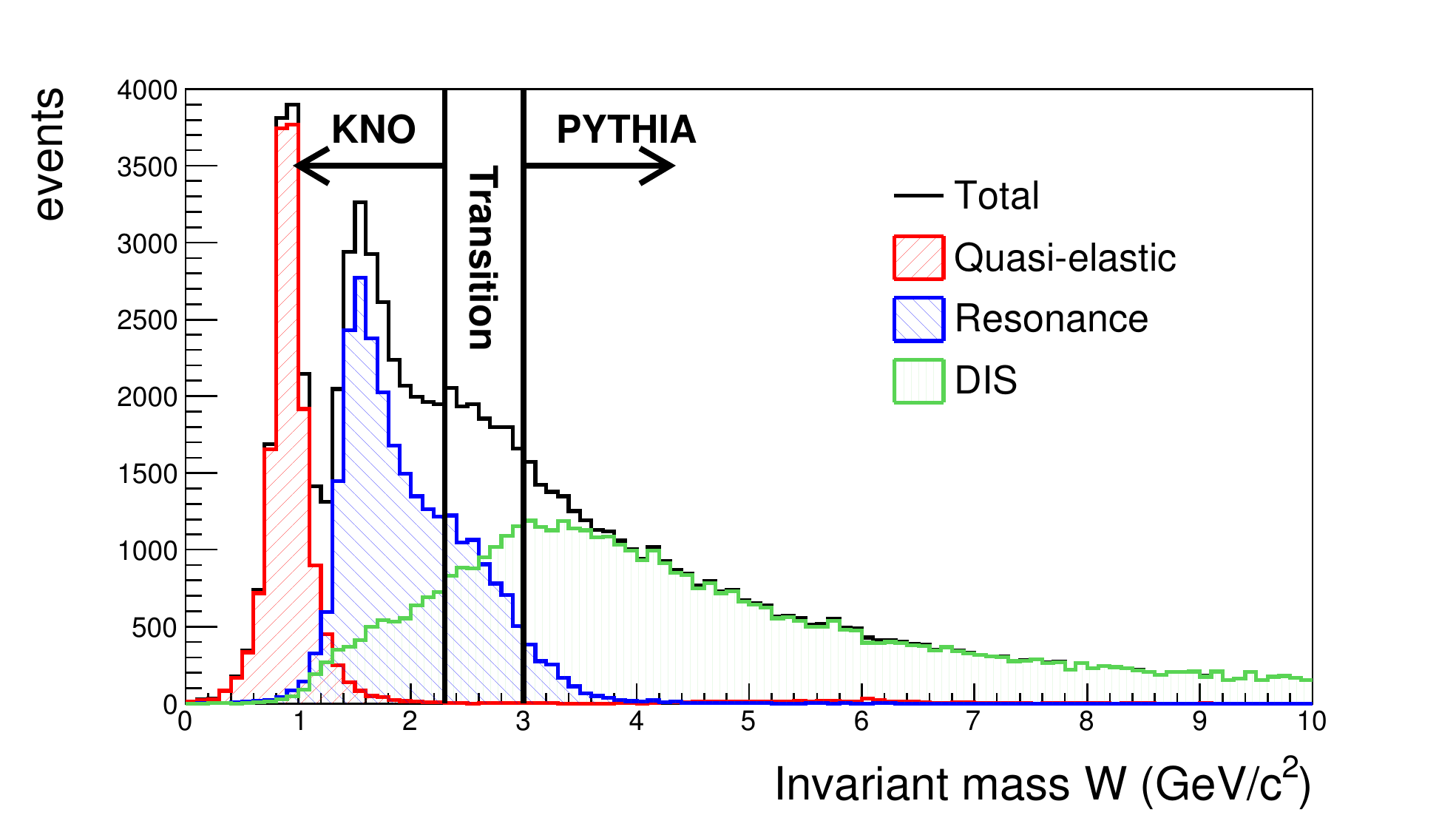}
\caption{\label{fig:KNOtoPYTHIA}
(color online) 
W distribution of $\numu$-water target interaction in GENIE. 
For the flux, we use an atmospheric $\numu$ neutrino spectrum. 
Left red hatched region is quasi-elastic scattering, 
middle hatched region is resonance interactions, 
and right green hatched region is from DIS.  
The $W$ distribution can be split to three regions, 
KNO scaling-based model only region, 
PYTHIA only region, and the transition region.
}
\end{figure}

Fig.~\ref{fig:KNOtoPYTHIA} graphically shows this situation. 
This is the $W$ distribution for $\numu$-water interactions simulated with GENIE. 
Here, we used a simple formula to model the atmospheric $\numu$ neutrino spectrum~\cite{Honda1,Honda2}, described later. 
As you can see, the $W$-distribution in this energy region can be split into three main interaction modes, 
quasi-elastic (red hatched, left peak), resonance (blue hatched, middle), 
and DIS (green hatched, right). The AGKY model is applied to DIS interactions.  
Also note DIS is extended to the low $W$ region to describe 
non-resonance interactions in the resonance region.
Although charm production processes are possible at the high energies,
contributions are minor and throughout this article we ignore charm production processes.

\vspace{0.5cm}
All studies in this paper use GENIE version 2.8.0, 
also Fig.s~\ref{fig:cMulCh}, \ref{fig:cMulCh_nubar}, \ref{fig:cXF}, \ref{fig:cMulPi0}, and \ref{fig:cTopo} are 
generated by the hadronization validation tool in GENIE.


\section{HERMES experiment\label{sec:hermes}} 

HERMES is a fixed target experiment at DESY~\cite{HERMES_MulCh}. 
The ring stores 27.6 GeV electrons or positrons, and collisions take place 
in the HERMES gas-jet target. 
  
The HERMES experiment has a long history of tuning PYTHIA for their purposes. 
The main motivation of this is because the default PYTHIA parameters are 
tuned to higher energy $e^+-e^-$ experiments ($\sqrt{s}\sim 35$~GeV)
and are not quite suitable for HERMES. 
Since modern neutrino oscillation experiments are also lower energy 
(1-10 GeV) compared with collider experiments, 
it is interesting to test the PYTHIA developed for the HERMES experiment within GENIE. 
Various alterations have been applied to PYTHIA by HERMES collaborators and among them, 
we are most interested in the adjustment to the fragmentation model made by tuning PYTHIA parameters,
without modifying the source code.

Parameter sets developed by HERMES collaborators are available elsewhere 
(for example, Ref.~\cite{Menden:2001gy,Rubin:2009zz,Hillenbrand:2005ke,HERMES_GluPol}). 
Table~\ref{tab:Parameter_values} summarizes the parameter sets we studied
(Lund-scan, $\De q(x)$, 2004c), as well as their default values in PYTHIA and GENIE.
In the second column, the default PYTHIA parameters are shown fully
while the second to the fifth column are shown only if the parameters
have been changed from the default PYTHIA values.  
In this article, we focus on the parameter set called ``Lund-scan''~\cite{Rubin:2009zz}, 
which we found had the best agreement with neutrino hadron production data. 

\begin{table}[!t]
    \centering
    \footnotesize
    \
    \begin{tabular}{c | c | c | c | c | c | l}
        Parameter & PYTHIA & GENIE & \texttt{Lund-scan} & $\De q(x)$&\texttt{2004c} & descriptions\\
        \hline \hline
        \texttt{PARJ1}  & 0.10 &      & 0.02 &      & 0.029 &di-quark suppression                     \\
        \texttt{PARJ2}  & 0.30 & 0.21 & 0.25 & 0.20 & 0.283 &strange quark suppression                \\
        \texttt{PARJ11} & 0.50 &      & 0.51 &      &       &light vector meson suppression           \\
        \texttt{PARJ12} & 0.60 &      & 0.57 &      &       &strange vector meson suppression         \\
        \texttt{PARJ21} & 0.36 & 0.44 & 0.42 & 0.37 & 0.38  &width of Gaussian $p_\perp$ distribution  \\
        \texttt{PARJ23} & 0.01 &      &      & 0.03 &       &non-Gaussian tail of $p_\perp$ distribution\\
        \texttt{PARJ33} & 0.80 & 0.20 & 0.47 &      &       &string breaking mass cutoff               \\
        \texttt{PARJ41} & 0.30 &      & 0.68 & 1.74 & 1.94  &Lund $a$ parameter                        \\
        \texttt{PARJ42} & 0.58 &      & 0.35 & 0.23 & 0.544 &Lund $b$ parameter                        \\
        \texttt{PARJ45} & 0.50 &      & 0.74 &      & 1.05  &adjustment of Lund $a$ for di-quark       \\
        \hline \hline
    \end{tabular}
    \caption{Parameter values for data sets \texttt{$ \Delta $q(x)}
        \cite{Menden:2001gy}, \texttt{Lund-scan} \cite{Rubin:2009zz} and
        \texttt{2004c} \cite{Hillenbrand:2005ke} along with the default
        PYTHIA and GENIE values.  Blank cells represent that the parameter
        takes on the default PYTHIA value.}
    \label{tab:Parameter_values}
\end{table}

Note we only tested PYTHIA parameters which are publicly available, 
however, HERMES also made modifications to the source code of PYTHIA itself. 
Therefore, in this paper we are not testing with the exact  
hadronization model used in the HERMES experiment. 
Also ``default GENIE'' quoted in this paper is not GENIE with default PYTHIA 6.3 parameters,
since by default, GENIE modified three parameters listed in Table~\ref{tab:Parameter_values},
however, we confirm the difference in predictions
between the default GENIE and the GENIE with the default PYTHIA parameters is very small within our studies. 

\section{Averaged charged hadron multiplicity\label{sec:avmulch}}

\begin{figure}[b]
\includegraphics[width=0.7\textheight]{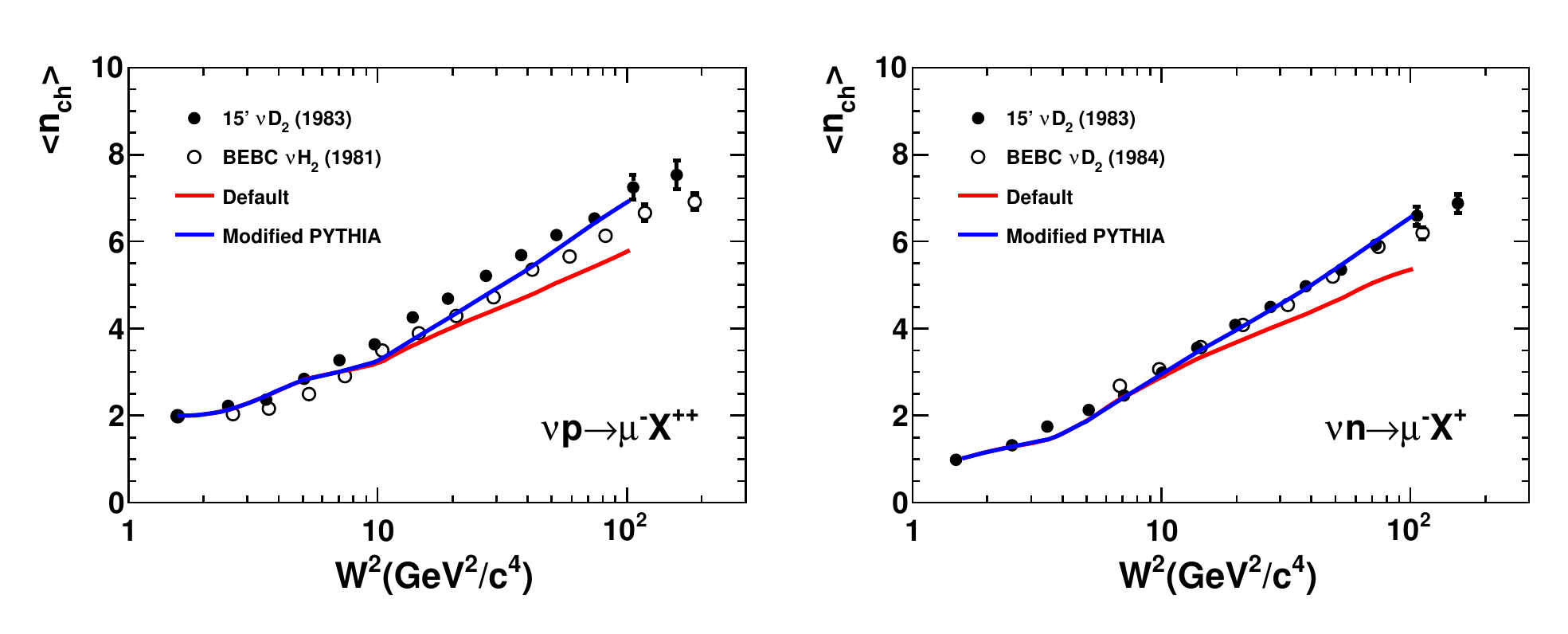}
\caption{\label{fig:cMulCh}
(color online) 
Averaged charged hadron multiplicity plot. 
Here, two predictions from GENIE are compared with bubble chamber $\numu-p$ 
and $\numu-n$ hadron production data~\cite{Zieminska,BEBC_D2}. 
}
\end{figure}

Averaged charged hadron multiplicity data is fundamental in the development of hadronization models. 
It describes the average number of charged hadrons, mainly $\pip$ and $\pim$, 
measured as a function of invariant hadron mass $W$. 
Neutrino hadronization models are largely guided by such data 
from bubble chamber experiments.   
Recently, Kuzmin and Naumov performed detailed surveys of neutrino bubble chamber data, 
and chose the best sets of data to tune their model~\cite{KuzminNaumov}. 
It is shown that all modern neutrino interaction generators, 
such as GENIE~\cite{GENIE}, NuWro~\cite{NuWro}, and GiBUU~\cite{GiBUU}, 
all appear to underestimate averaged charged hadron multiplicity.
Note, it is also shown that the NEUT neutrino interaction generator~\cite{NEUT},
which is used by T2K and Super-Kamiokande, 
also underestimates averaged charged hadron multiplicity~\cite{Connolly}. 

This problem largely originates from the PYTHIA fragmentation model, 
because as mentioned in the previous section, 
the default PYTHIA parameters are tuned to higher energy experiments. 
Both GENIE and NuWro~\cite{Nowak_pythia} tuned these PYTHIA parameters to improve 
the agreement with data but the effect is marginal. 
Note NuWro and GiBUU use their own models for fragmentation, 
and only later processes are based on PYTHIA. 

Fig.~\ref{fig:cMulCh} shows the data-MC comparison of the averaged charged hadron multiplicity 
in $\numu-p$ and $\numu-n$ interactions. 
Here, the two curves represent predictions from default GENIE and 
GENIE with a PYTHIA modified using the Lund-scan parameter set~\cite{Rubin:2009zz}. 
Note, because $W<2.3~\uGeVct$ range of the AGKY model is hadronized using the KNO scaling-based model,
these two curves should be identical at $W<2.3~\uGeVct$. 
As you can see, the HERMES tune describes the data better. 
Here, two data sets from the Fermilab bubble chamber and BEBC agree in $\numu-n$ interactions (both deuterium targets) 
but not in $\numu-p$ data (hydrogen and deuterium target),  
suggesting the conflict of data we see in Fig.~\ref{fig:cMulCh} is 
due to nuclear effects in deuterium~\cite{AGKY,Tingjun,KuzminNaumov,Nowak_had}. 
Despite the conflict of data,
the HERMES parameterization in general increases the averaged charged hadron multiplicity, 
which improves the agreement with averaged charged hadron multiplicity data from neutrino bubble chamber experiments. 

\begin{figure}[tb]
\includegraphics[width=0.5\textheight]{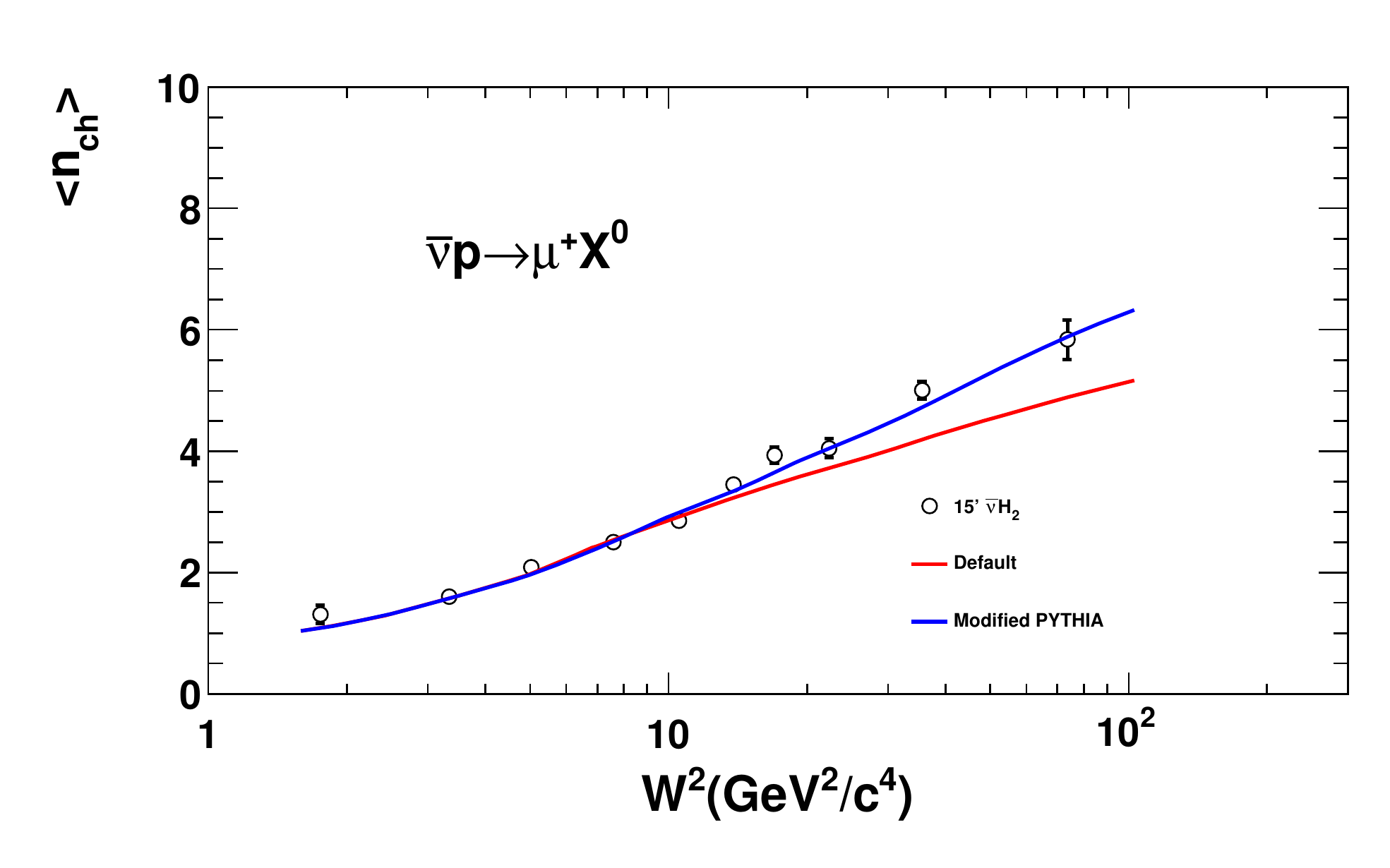}
\caption{\label{fig:cMulCh_nubar}
(color online) 
Averaged charged hadron multiplicity plot. 
Here, two predictions from GENIE are compared with bubble chamber 
$\numubar-p$ hadron production data~\cite{Derrick}. 
}
\end{figure}

Fig.~\ref{fig:cMulCh_nubar} is the same plot for $\numubar-p$ interactions. 
Again, the agreement with the data is better for GENIE with the modified PYTHIA parameter set. 
Therefore, this new parameter set works better for both neutrino and antineutrino interactions.

The main effect of this new parameterization originates from 
the increase of the Lund $a$ parameter and the decrease of the Lund $b$ parameter (Eq.~\ref{eq:lundff}). 
As shown in Fig.~\ref{fig:lundfunc},
these changes make the fragmentation function softer. 
This enhances emissions of soft hadrons, {\it i.e.},
this increases averaged charged hadron multiplicity and thus it agrees better with data. 
In the higher energy experiments that PYTHIA is designed for, 
high order QCD effects cause additional low energy parton emissions. 
This causes hadrons to be produced with a broader spectrum in $z$. 
Because these effects are negligible for the neutrino experiments we are concerned with, 
we shift the peak of the fragmentation function to a lower $z$ value by increasing (decreasing) 
the Lund $a$ ($b$) parameter~\cite{Hillenbrand:2005ke}.

In fact, all parameterization schemes from HERMES we checked (Table~\ref{tab:Parameter_values})
have a high (low) Lund $a$ ($b$) parameter,
and many have more extreme values than the ones we use here. 
However, these higher (lower) Lund $a$ ($b$) parameter models tend to overestimate hadron multiplicities 
compared to neutrino hadron production data from bubble chamber experiments 
and as a result the data-MC agreement becomes worse. 
The neutrino hadronization data prefer a relatively smaller Lund $a$ parameter than most HERMES parameter sets, 
yet bigger than the default PYTHIA choice, 
and this is the main reason why we chose this specific parameterization scheme in this paper. 

\section{$x_F$ distribution\label{sec:xf}}

\begin{figure}[tb]
\includegraphics[width=0.7\textheight]{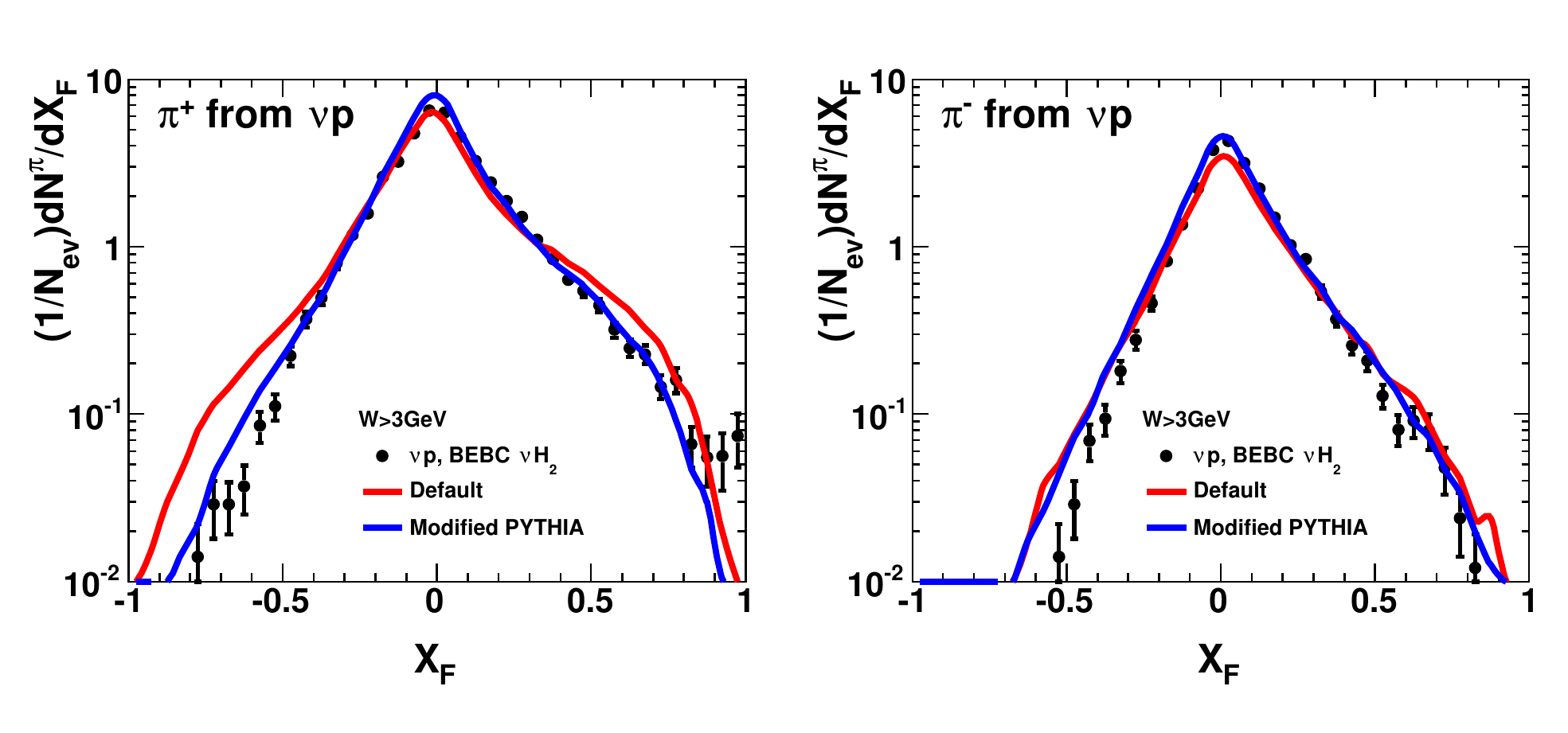}
\caption{\label{fig:cXF}
(color online)
$x_F$ distribution for $\pip$ and $\pim$ from $\numu-p$ interactions~\cite{BEBC_xF}. 
Again, modified PYTHIA has a better agreement with data.
}
\end{figure}

Feynman x, $x_F$, is the fraction of longitudinal momentum available for a hadron, 
defined in the hadronic center mass system,
{\it i.e.}, $x_F=\frac{P_L^*}{P_{Lmax}^*}\sim\frac{2P_L^*}{W}$ , 
here asterisks stand for the hadron c.m.s. 
Fig.~\ref{fig:cXF} shows the data-MC comparison. 
The agreement of modified PYTHIA with bubble chamber data is excellent 
for both $\pip$ and $\pim$ data. 
Therefore the tuning we applied is valid not only for averaged charged hadrons, 
but also valid for positive and negative hadrons separately.

\section{Averaged neutral pion multiplicity\label{sec:avmulpiz}}

\begin{figure}[t]
\includegraphics[width=.35\textheight]{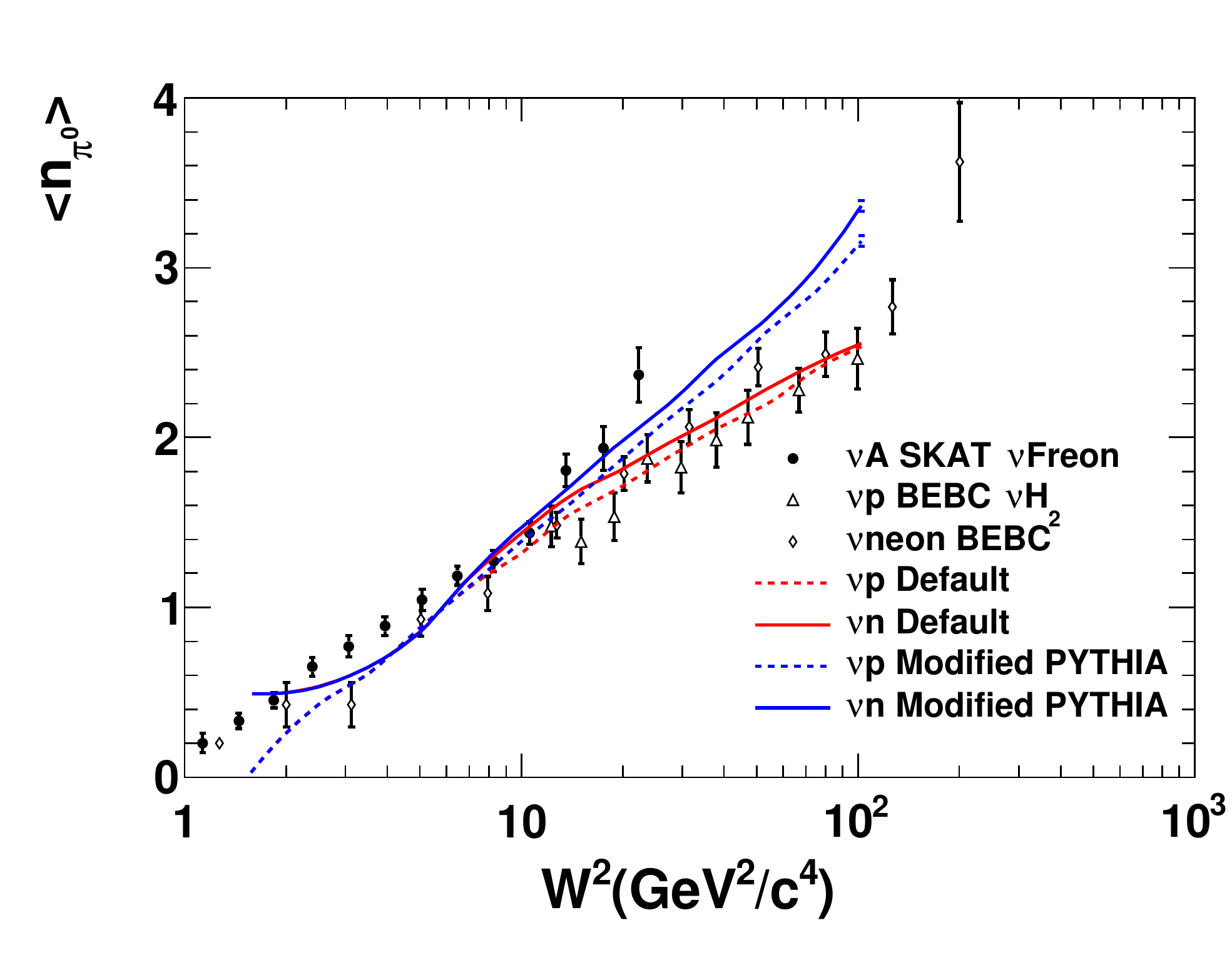}
\includegraphics[width=.35\textheight]{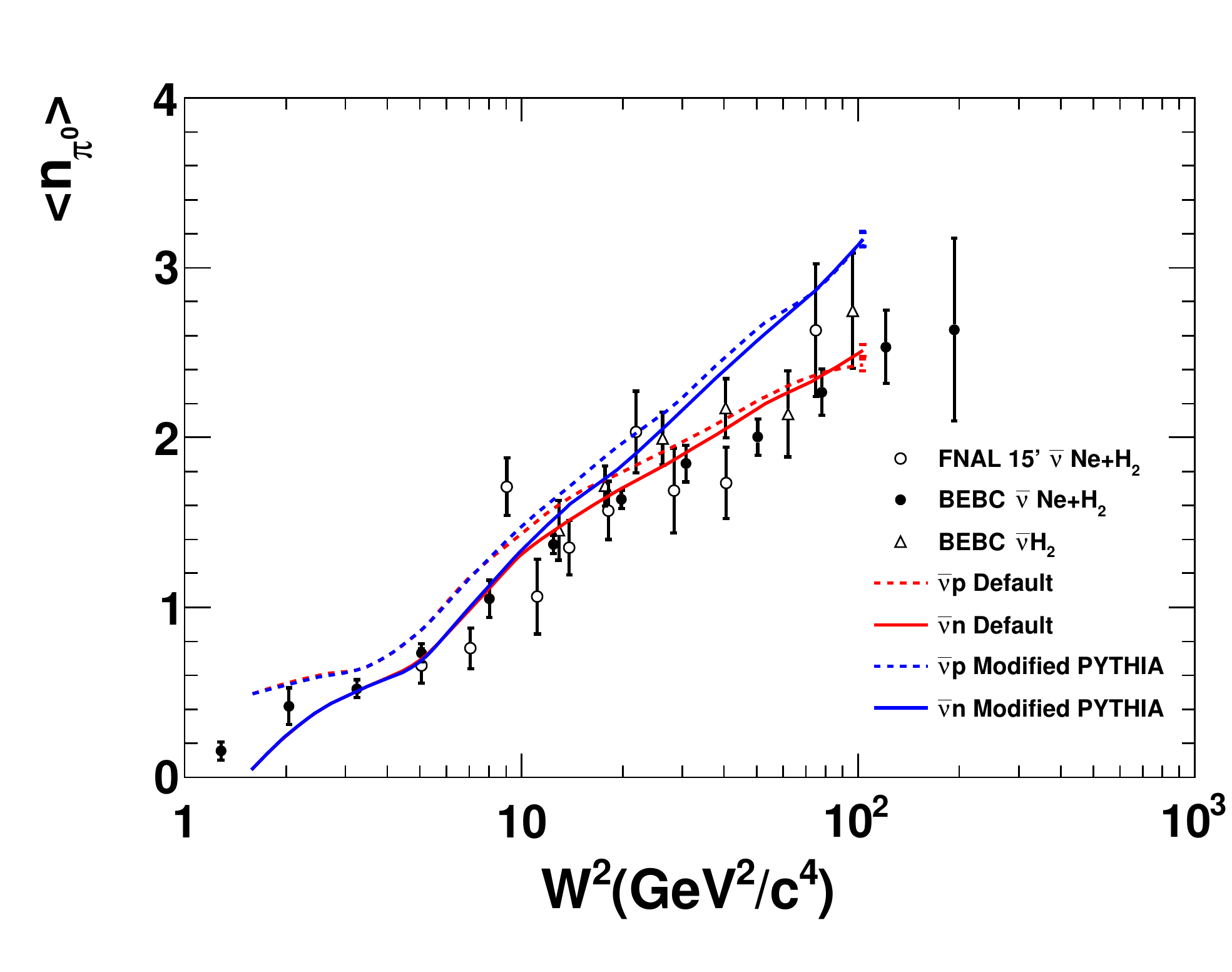}
\caption{\label{fig:cMulPi0}
(color online) 
Averaged neutral pion multiplicity plot. 
Here, two predictions from GENIE are compared with bubble chamber $\numu-p$, $\numu-n$,  
$\numubar-p$, and $\numubar-n$ $\piz$ production data~\cite{BEBC_Ne,SKAT_F,BEBC_H2,NC51A}.
}
\end{figure}

In Fig.~\ref{fig:cMulPi0}, predictions are compared with the averaged $\piz$ multiplicity. 
Here the data from $\numu$ and $\numubar$ interactions are from various targets~\cite{BEBC_Ne,SKAT_F,BEBC_H2,NC51A}.
Although the data here have larger errors, 
now the default GENIE has a better agreement with the data. 
The ratio of the number of produced charged pions and neutral pions is strongly tied due to isospin symmetry, 
{\it i.e.} $N(\pip)+N(\pim):N(\piz)=2:1$. 
Thus, if we increase the charged hadron multiplicity, 
the model will also have higher multiplicities of neutral pions.  
The charged pion and neutral pion multiplicity ratio is 2:1 in BEBC neon target bubble chamber data~\cite{BEBC_Ne},   
however, this relationship is not obvious in other bubble chamber data.
As we see from Fig.~\ref{fig:cMulCh} and~\ref{fig:cMulPi0}, 
it is not easy to achieve good agreements with both charged hadron and neutral pion multiplicities 
by tuning PYTHIA parameters. 
On the other hand, PYTHIA shows excellent agreements in both charged and neutral pion fragmentation functions 
with HERMES data~\cite{HERMES_MulCh,Joosten}.


\section{Topological cross sections\label{sec:topo}}


In the low W region, PYTHIA does not predict the multiplicity properly. 
In GENIE, the AGKY model uses a phenomenological approach based on KNO scaling~\cite{KNO}, 
where dispersion is assumed to follow a scaling law as data suggest.
Thus, by definition, the AGKY model has a better data-MC agreement for the dispersion of the multiplicity in the low W region. 
This is not the case in PYTHIA, where physics is simulated from a more first principles approach,  
which is based on quark-diquark fragmentation.
By tuning PYTHIA parameters, the data-MC agreement of 
the averaged charged hadron multiplicity can be improved, 
but it is not easy to fully correct the dispersion. 

\begin{figure}[tb]
\includegraphics[width=.35\textheight]{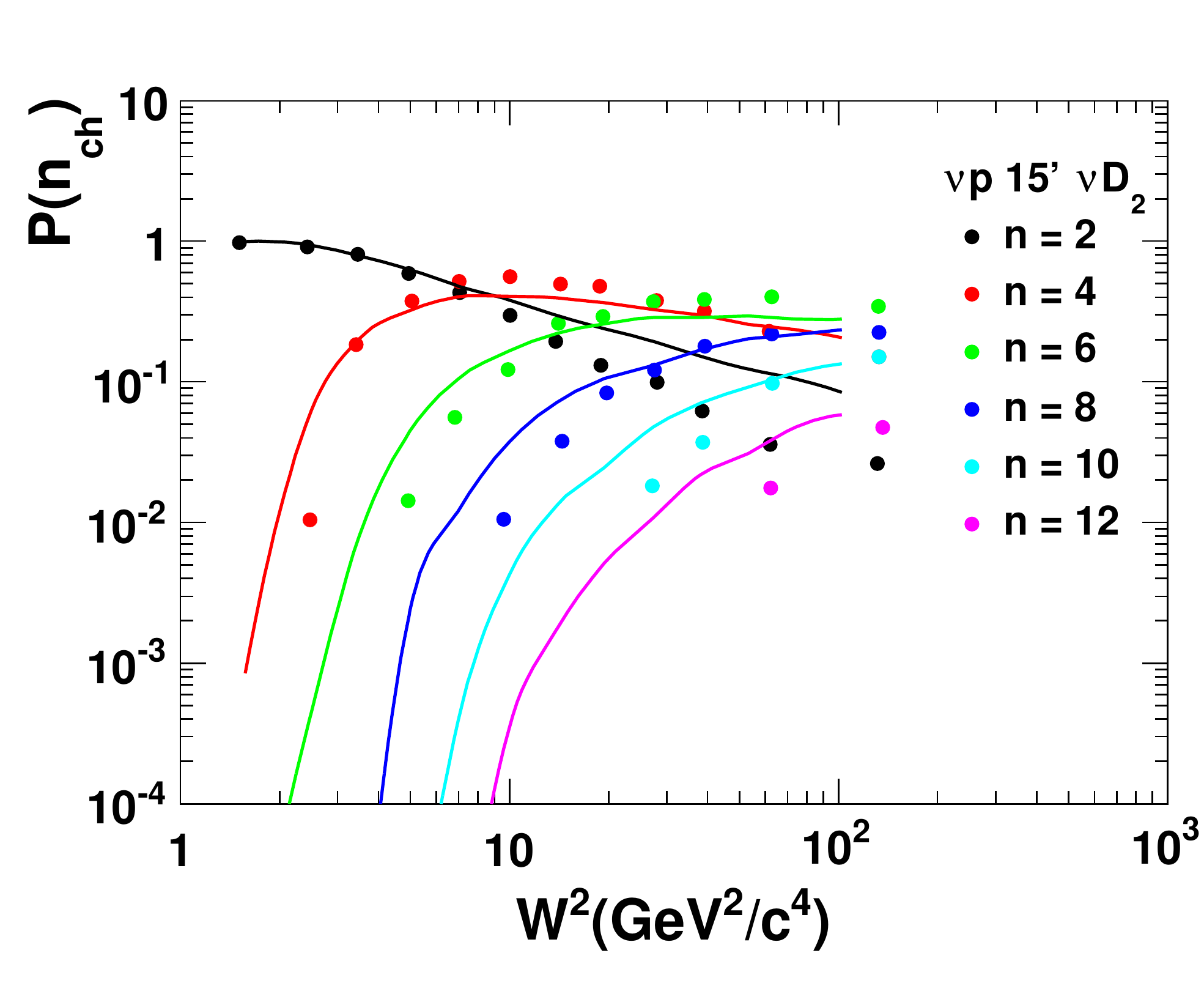}
\includegraphics[width=.35\textheight]{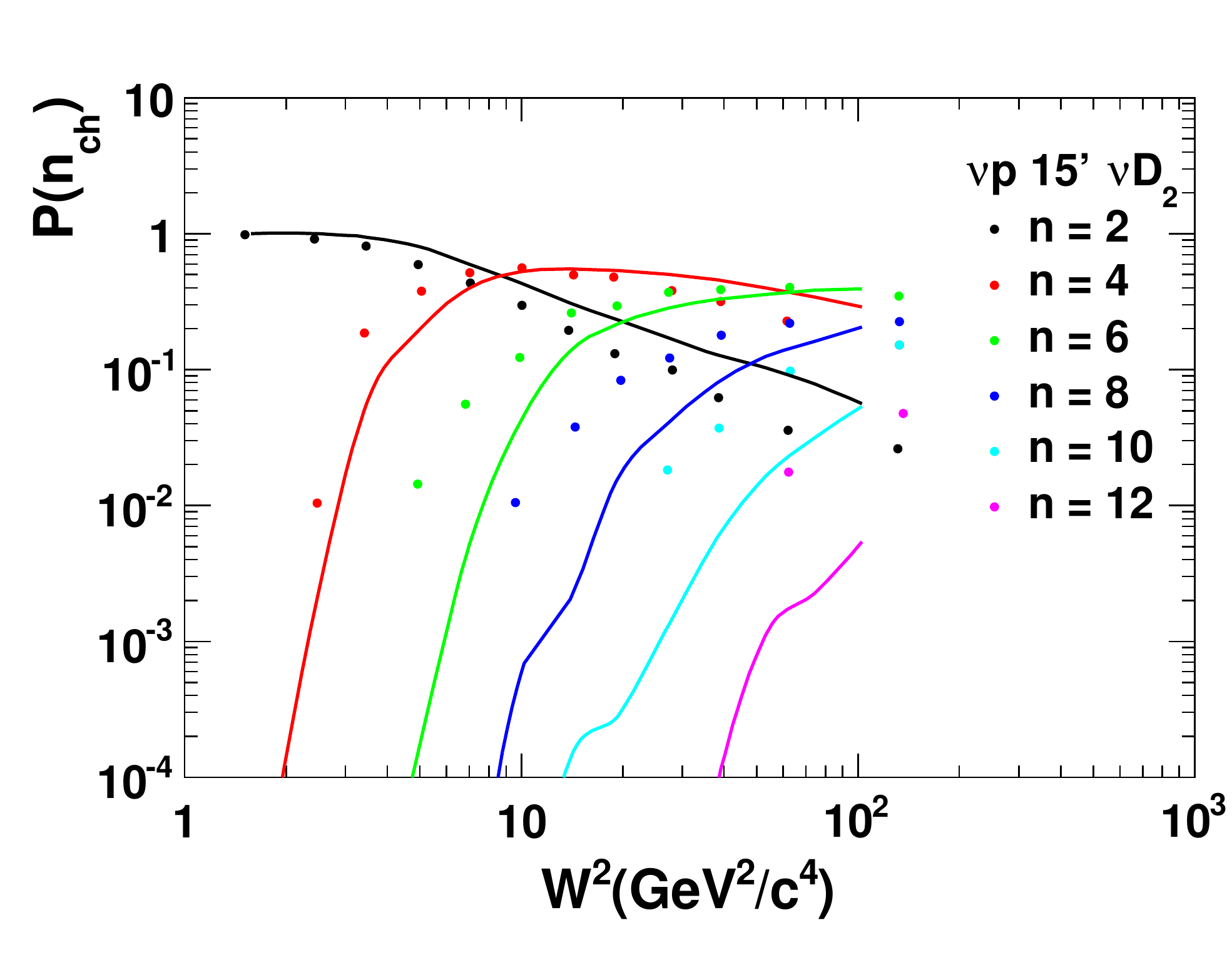}
\caption{\label{fig:cTopo}
(color online)
Topological cross sections of charged hadrons for $\numu-p$ interaction. 
In both plots, data points are from deuteron target Fermilab 15' bubble chamber experiments~\cite{Zieminska}. 
In the left plot, PYTHIA is turned off and data are compared with GENIE 
with only the KNO scaling-based hadronization model. 
On the other hand, in the right plot, hadronization is solely handled by PYTHIA.
}
\end{figure}

Fig.~\ref{fig:cTopo} shows the data-MC comparisons of the topological cross sections of charged hadrons, that is, 
the fraction of final particle topologies of a given interaction as function of invariant mass. 
In both plots, the GENIE predictions are compared with deuteron target $\numu-p$ data
from Fermilab 15' bubble chamber~\cite{Zieminska}. 
In the left plot, the hadronization model in GENIE is solely carried out by the KNO scaling-based model. 
Since the KNO scaling-based approach is designed to reproduce the dispersion data, 
GENIE can make the large multiplicities, such as n=6, n=8, etc, as data suggested. 
On the other hand, in the right plot, 
the GENIE hadronization model solely depends on PYTHIA. 
In this case, we see PYTHIA has problems reproducing large hadron multiplicity events. 
Therefore, the combination of KNO scaling-based model and PYTHIA cannot make smooth topological cross sections.
This indicates the dispersion of hadron multiplicity reproduced by PYTHIA is smaller than the data. 

High resolution liquid argon time projection chamber (LArTPC) experiments, 
such as MicroBooNE~\cite{Katori:2011uq}, are in a good position to identify high multiplicity hadron events. 
These data may offer the opportunity to test neutrino hadronization processes.   
However, to test hadronization models with hadron data from heavy nuclear targets such as argon, 
it is also necessary to have a good model for primary interactions~\cite{Morfin,Hayato,Zeller_INT} 
and nuclear effects~\cite{Mosel_had}. 
In inelastic interaction processes, both primary interactions and nuclear effects play significant roles and 
currently disagreements between data and predictions are not well understood~\cite{Sobczyk_SPP}.  
Therefore, it is challenging to develop a hadronization model solely from neutrino experimental data, 
and input from other fields, especially electron scattering experiments, is very important.

\section{Impact on hadronization models for PINGU\label{sec:pingu}}

\begin{figure}[b]
\includegraphics[width=.6\textheight]{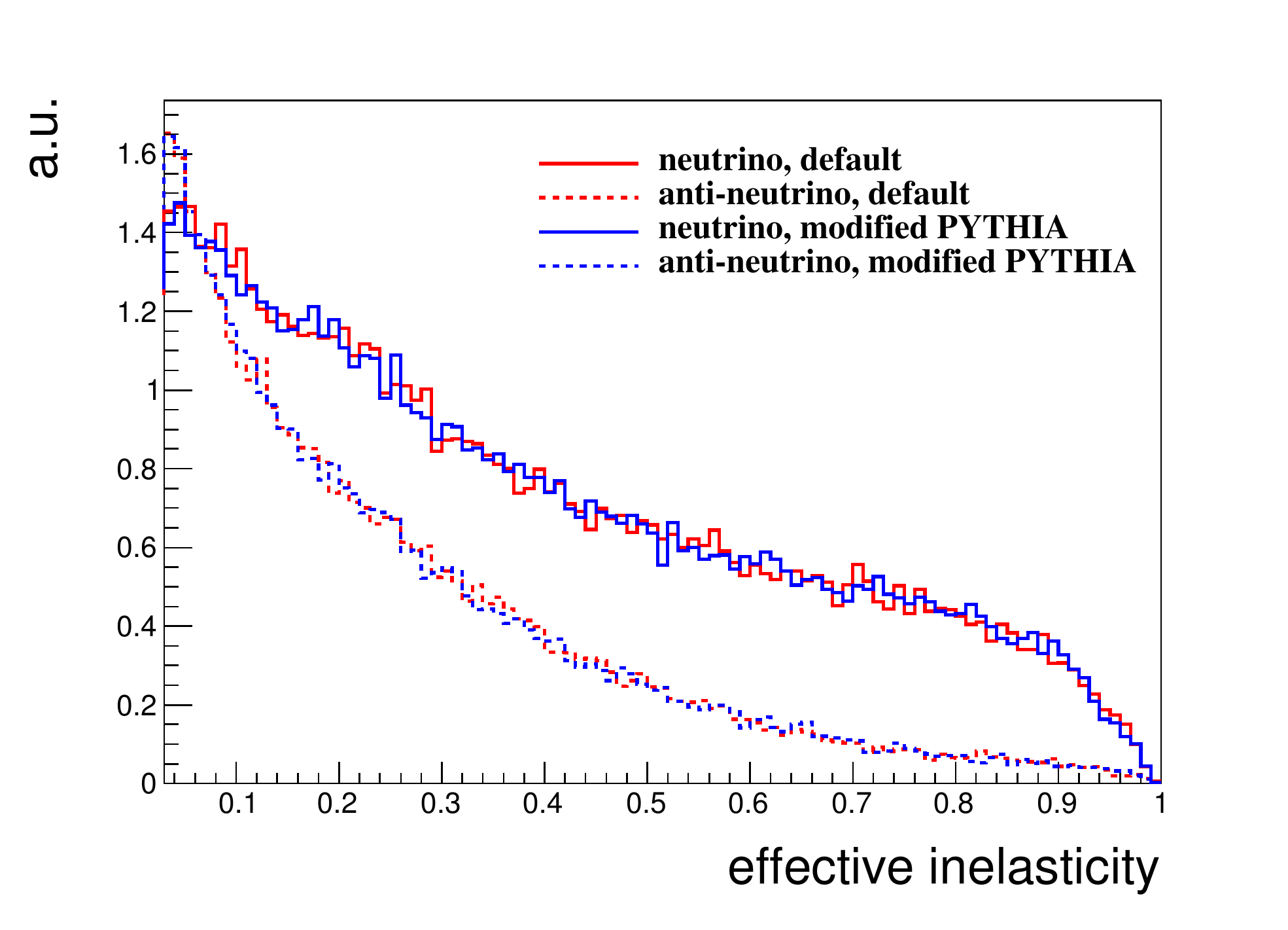}
\caption{\label{fig:ydistri}
(color online)
Effective inelasticity distribution with atmospheric neutrino spectrum. 
Here, all histograms are arbitrarily normalized. 
Solid histograms are muon neutrino distributions, and dashed histograms are muon anti-neutrino distributions. 
Red histograms are from GENIE with the default hadronization model, 
and blue histograms are from GENIE with the modified hadronization model discussed in this paper. 
}
\end{figure}

In previous sections, we discussed how PYTHIA can be improved
inside GENIE in order to reproduce neutrino bubble chamber data. 
However, it is not obvious how such improvements affect current and future neutrino experiments, 
unless a realistic neutrino flux is integrated in the interaction simulations.
In this section, we use an atmospheric neutrino flux prediction 
to simulate neutrino interactions in order to
study how different hadronization models affect analysis in PINGU~\cite{PINGU},
which will try to use hadronic information to improve their sensitivity on oscillation physics. 

PINGU is a low energy extension of the IceCube detector~\cite{IceCube}. 
By placing optical sensors closer together compared to the original IceCube detector, 
PINGU is able to measure atmospheric neutrinos below 20 GeV where matter oscillations are important. 
Although PINGU has a significantly smaller volume coverage compared with the 1km$^3$ IceCube detector, 
the estimated PINGU volume coverage is still $\sim$6~Mton and high statistics is expected.   
The capability of atmospheric neutrino oscillation measurements has also been demonstrated recently~\cite{DeepCore}.

The goal of PINGU is to determine the neutrino mass hierarchy (NMH) through matter oscillations. 
In the two-neutrino oscillation approximation, 
the muon neutrino disappearance oscillation probability in the normal hierarchy ($P_{\al\be}^{NH}$) 
and the muon anti-neutrino disappearance oscillation probability in the inverted hierarchy ($\bar{P}_{\al\be}^{IH}$) 
are the same ($P_{\al\be}^{NH}=\bar{P}_{\al\be}^{IH},~\bar{P}_{\al\be}^{NH}=P_{\al\be}^{IH}$)~\cite{PINGU_inel}. 
Thus, it is also desirable to separate muon neutrinos and muon anti-neutrinos where final state leptons are indistinguishable 
by Cherenkov detectors such as PINGU.  

Recently, Ribordy and Smirnov pointed out that the charge separation,    
through the precise measurement of inelasticity distributions, 
improves the PINGU and ORCA NMH sensitivity~\cite{PINGU_inel}. 
The same arguments may be applied to Hyper-Kamiokande~\cite{HyperK} and DUNE~\cite{LBNE}. 
Since inelasticity measurements rely on the energy deposits of hadronic showers, 
it is interesting to check the impact of different hadronization models in this situation. 

For this purpose, we estimated the impact of hadronization models on the effective inelasticity. 
We define the effective inelasticity from the visible hadron shower energy. 
\beq
E_h^{vis}=\sum_{E_h^i>E_{th}^i}T_h^i+\sum E_\ga^i~.
\eeq
The first term is the sum of kinetic energies of charged hadrons above the Cherenkov threshold.  
Here we assume that the charged hadrons above the Cherenkov threshold are
visible and we take into account their kinetic energies. 
The second term is the sum of all the final state photons, including the decays of neutral mesons. 
Thus, the visible hadron energy corresponds to the energy deposit from the hadronic system to the perfect photon detector, 
where inefficiency is only from neutrons or hadrons below the Cherenkov threshold. 
Then, the effective inelasticity, $y^{eff}$, is defined by,
\beq
y^{eff}=\frac{E_h^{vis}}{E_h^{vis}+E_\mu}~.
\eeq
Here $E_\mu$ is the muon (anti-muon) energy.

To simulate effective inelasticity on a water target, 
we modeled the atmospheric neutrino flux with a simple formula ($\sim a+b\cdot E^{-c}$, where $c\sim 2.8$) 
which reproduces the typical energy spectrum of atmospheric neutrinos~\cite{Honda1,Honda2}. 
Then we simulate neutrino interactions from 2 to 30~GeV,
which is the energy region of interest for NMH analysis. 
 
Fig.~\ref{fig:ydistri} shows the simulated $y^{eff}$ distributions with arbitrary normalization. 
The $y^{eff}$ distributions for neutrino and anti-neutrino interactions are well separated, 
however, $y^{eff}$ distributions based on different hadronization models are very similar.
This study does not include the detector simulation,
and the quantitative evaluation of the impact of hadronization models in PINGU is difficult. 
However, the inclusion of a detector simulation in general smears structures and
it will make the results of two different hadronization models nearly identical. 
This result can be understood from a simulated $W$ distribution of PINGU.
PINGU is dominated by few GeV low $W$ interactions ($<5~GeV$) 
where PYTHIA hadronization processes have a minor role (Fig.~\ref{fig:KNOtoPYTHIA}). 
This indicates alternations of the hadronization model only provide 
minor changes to the systematics of the PINGU inelasticity measurement, 
however, details have yet to be tested with a full PINGU detector simulation. 

\section*{Conclusion\label{sec:conclusion}}

In this article, we studied neutrino hadronization processes in GENIE. 
Our main focus is to improve the averaged charged hadron multiplicity, 
and it is shown that suitable parameterization developed by the HERMES collaboration dramatically 
improves the data-MC agreement with neutrino bubble chamber data. 
However, this tuning may make the $\piz$ multiplicity agreement slightly worse. 
Also dispersion of hadron multiplicity is still not under control. 
All studies in the paper are qualitative in nature,
and quantitative studies are beyond the scope of this paper.
At the present moment, PYTHIA parameter tunings based on fits to
neutrino experimental data are not being performed. 
An example of the difficulty of PYTHIA tunings
is the correlations between different PYTHIA parameters,
as HERMES collaborators noted~\cite{Rubin:2009zz}.
In this article, we find that controlling the shape of the Lund string fragmentation function 
using the Lund $a$ and Lund $b$ parameters is the key to control the averaged hadron multiplicity, 
and a more precise tuning is left as a direction for future work.
Near future LArTPC experiment, such as MicroBooNE, 
could test the hadronization models by measuring high hadron multiplicity events.  

Finally, we tested different hadronization models with a modeled atmospheric neutrino flux. 
It is seen that the difference in the inelasticity distributions is small, 
suggesting hadronization processes only play a minor role in the systematics for NMH analysis 
at atmospheric neutrino oscillation experiments.
Careful analysis including the detector simulation will find more accurate systematics of
hadronization interactions for future atmospheric neutrino experiments,
such as PINGU, ORCA, Hyper-Kamiokande and DUNE. 


\section*{Acknowledgment}
TK thanks Ulrich Mosel for introducing this subject to us. 
We thank Elke Aschenauer and Josh Rubin for useful information about the HERMES experiment. 
We also appreciate the various help given to us by Gabe Perdue and Julia Yarba on the GENIE simulations. 
Finally, TK would like to thank the organizer of CETUP* 
(Center for Theoretical Underground Physics and Related Areas) 
for the hospitality during my stay at Deadwood, SD to engage on this work.

\bibliographystyle{aipproc}   

\bibliography{hermespythia6}
\end{document}